\documentclass[]{spie}

\usepackage{amsmath,amsfonts,amssymb}
\usepackage{graphicx}
\usepackage[colorlinks=true, allcolors=blue]{hyperref}
\usepackage{orcidlink}
\usepackage{multirow}
\usepackage{makecell}
\usepackage{colortbl}
\usepackage{amsmath}

\title{Calibration Measurements of the BICEP3 and BICEP Array CMB Polarimeters from 2017 to 2024}

\author[a]{C. Giannakopoulos}
\author[b]{C. Vergès}
\author[c]{P.A.R. Ade}
\author[d]{Z. Ahmed}
\author[e]{M. Amiri}
\author[b]{D. Barkats}
\author[f]{R. Basu Thakur}
\author[a]{C.A. Bischoff}
\author[d,g]{D. Beck}
\author[f,i]{J.J. Bock}
\author[b]{H. Boenish}
\author[j]{V. Buza}
\author[k]{J.R. Cheshire IV}
\author[b]{J. Connors}
\author[b]{J. Cornelison}
\author[k]{M. Crumrine}
\author[f]{A.J. Cukierman}
\author[l]{E.V. Denison}
\author[b]{M.I. Dierickx}
\author[n]{L. Duband}
\author[b]{M. Eiben}
\author[b,q]{B.~D.~Elwood \orcidlink{0000-0003-4117-6822}}
\author[f]{S. Fatigoni}
\author[o,p]{J.P. Filippini}
\author[g]{A. Fortes}
\author[f]{M. Gao}
\author[g]{ N. Goeckner-Wald}
\author[d]{D.C. Goldfinger}
\author[g]{J. Grayson}
\author[b]{P.K. Grimes}
\author[k]{G. Hall}
\author[g]{G. Halal}
\author[e]{M. Halpern}
\author[a]{E. Hand}
\author[b]{S.A. Harrison}
\author[d]{S. Henderson}
\author[l]{J. Hubmayr}
\author[f]{H. Hui}
\author[l,d,g]{K.D. Irwin}
\author[g,f]{J. Kang}
\author[d]{K.S. Karkare}
\author[f]{S. Kefeli}
\author[b]{J.M. Kovac}
\author[g,d]{C.L. Kuo}
\author[f]{K. Lau}
\author[a]{M. Lautzenhiser}
\author[o]{A. Lennox}
\author[g]{T. Liu}
\author[i]{K.G. Megerian}
\author[f]{O. Miller}
\author[f]{L. Minutolo}
\author[f]{L. Moncelsi}
\author[g]{Y. Nakato}
\author[i]{H.T. Nguyen}
\author[g,l]{R. O’Brient}
\author[f]{A. Patel}
\author[b]{M.A. Petroff}
\author[b]{A. Polish}
\author[k]{N. Precup}
\author[n]{T. Prouve}
\author[k,m]{C. Pryke}
\author[l]{C.D. Reinsema}
\author[f]{T. Romand}
\author[g]{M. Salatino}
\author[f]{A. Schillaci}
\author[b]{B.L. Schmitt}
\author[k]{B. Singari}
\author[f,i]{A. Soliman}
\author[b]{T. St. Germaine}
\author[f]{A. Steiger}
\author[f]{B. Steinbach}
\author[c]{R.V. Sudiwala}
\author[g,d]{K.L. Thompson}
\author[b]{C. Tsai}
\author[c]{C. Tucker}
\author[i]{A.D. Turner}
\author[j,h]{A.G. Vieregg}
\author[f]{A. Wandui}
\author[i]{A.C. Weber}
\author[k]{J. Willmert}
\author[d]{W.L.K. Wu}
\author[g]{H. Yang}
\author[g]{C. Yu}
\author[b]{L. Zeng}
\author[g,f]{C. Zhang}
\author[f]{S. Zhang}

\affil[a]{Department of Physics, University of Cincinnati, Cincinnati, Ohio 45221, USA}
\affil[b]{Center for Astrophysics \textbar~Harvard \& Smithsonian, Cambridge, Massachusetts 02138, U.S.A}
\affil[c]{School of Physics and Astronomy, Cardiff University, Cardiff, CF24 3AA, United Kingdom}
\affil[d]{Kavli Institute for Particle Astrophysics and Cosmology, SLAC National Accelerator Laboratory, Menlo Park, California 94025, USA}
\affil[e]{Department of Physics and Astronomy, University of British Columbia, Vancouver, British Columbia V6T 1Z1, Canada}
\affil[f]{Department of Physics, California Institute of Technology, Pasadena, California 91125, USA}
\affil[g]{Department of Physics, Stanford University, Stanford, California 94305, USA}
\affil[h]{Department of Physics, Enrico Fermi Institute, University of Chicago, Chicago, Illinois 60637, USA}
\affil[i]{Jet Propulsion Laboratory, Pasadena, California 91109, USA}
\affil[j]{Kavli Institute for Cosmological Physics, University of Chicago, Chicago, Illinois 60637, USA}
\affil[k]{Minnesota Institute for Astrophysics, University of Minnesota, Minneapolis, Minnesota, 55455, USA}
\affil[l]{National Institute of Standards and Technology, Boulder, Colorado 80305, USA}
\affil[m]{School of Physics and Astronomy, University of Minnesota, Minneapolis, Minnesota 55455, USA}
\affil[n]{Service des Basses Températures, Commissariat à l’Energie Atomique, 38054 Grenoble, France}
\affil[o]{Department of Physics, University of Illinois at Urbana-Champaign, Urbana, Illinois 61801, USA}
\affil[p]{Department of Astronomy, University of Illinois at Urbana-Champaign, Urbana, Illinois 61801, USA}
\affil[q]{Department of Physics, Harvard University, Cambridge, MA 02138, USA}

\authorinfo{Corresponding Author: Christos Giannakopoulos \\ 
E-mail: giannacs@mail.uc.edu}

\begin{document} 
\maketitle

\begin{abstract}
The BICEP3 and BICEP Array polarimeters are small-aperture refracting telescopes located at the South Pole designed to measure primordial gravitational wave signatures in the Cosmic Microwave Background (CMB) polarization, predicted by inflation. Constraining the inflationary signal requires not only excellent sensitivity, but also careful control of instrumental systematics.
Both instruments use  antenna-coupled orthogonally polarized detector pairs, and the polarized sky signal is reconstructed by taking the difference in each detector pair. As a result, the differential response between detectors within a pair becomes an important systematic effect we must control. Additionally, mapping the intensity and polarization response in regions away from the main beam  can inform how sidelobe levels affect CMB measurements. Extensive calibration measurements are taken \textit{in situ} every austral summer for control of instrumental systematics and instrument characterisation. In this work, we detail the set of beam calibration measurements that we conduct on the BICEP receivers, from deep measurements of main beam response to polarized beam response and sidelobe mapping. We discuss the impact of these measurements for instrumental systematics studies and design choices for future CMB receivers.

\end{abstract}


\section{INTRODUCTION}
\label{sec:into}

The Cosmic Microwave Background offers a unique window on the early Universe. 
Emitted just 380,000 years after the Big Bang at a time when photons were able to travel freely for the first time, the temperature and polarization of this first light are a prime source of information to constrain cosmological models.

The CMB has a uniform average temperature of
2.725K,\cite{bennett1993scientific} and fluctuations in CMB temperature at the level of $\sim100\mu$K~\cite{collaboration2013planck} probe the density fluctuations in the early Universe. 
The polarized CMB signal can be divided into curl-free fluctuations ($E$ modes) that also probe density fluctuations at the level of $\sim5\mu$K\cite{kovac2002detection}, and divergence-free fluctuations ($B$ modes) that probe tensor fluctuations in the early Universe.
Tensor fluctuations seeded by inflation\cite{guth1982fluctuations}~are much fainter than $E$ modes, and their amplitude is parameterized by the tensor-to-scalar ratio $r$.
Detection of excess power in the CMB polarization spectrum in the form of $B$~modes would constitute strong observational evidence of inflation. 
The most sensitive constraint to date is $r < 0.036$ at $95\%$ confidence level with $\sigma(r) = 0.009$, set using data from the BICEP/\textit{Keck} series of experiment up to the 2018 observing season (hereafter refered to BK18)~\cite{ade2021improved}.

\begin{figure}[ht!]
\centering
\includegraphics[width=15cm]{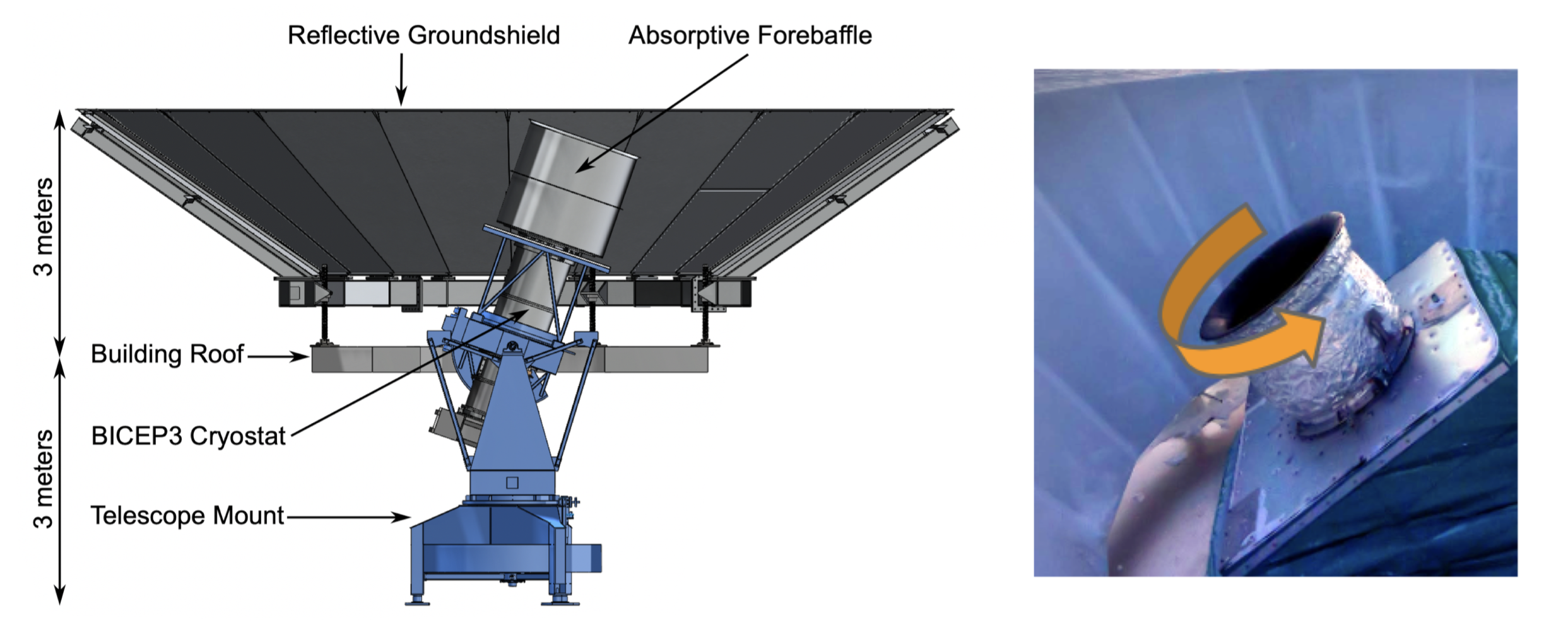}
\caption{BICEP3 mount, illustrating three degrees of freedom of Azimuth, Elevation, and Boresight (right). Various mount elements are also labeled. The cylindrical, co-moving, and blackened forebaffle extends above the cryostat to reduce sidelobe pickup. In addition, the mount is surrounded by a stationary reflective ground shield to reflect any off-field of view light onto the sky. Image courtesy of BICEP/\textit{Keck} Collaboration adapted from Figure 1 of previous publication\cite{boenishbicep}.}
\label{fig:boresight}
\end{figure}

Constraining $r$ is extremely difficult due to how faint $B$-mode fluctuations are compared to $E$ modes and other sources of celestial polarization like CMB lensing and Galactic foregrounds.
The BICEP3 \cite{ade2022bicep} and BICEP Array \cite{hui2018bicep} (BA) polarimeters located at the South Pole are specifically designed to measure the B-mode signature in the CMB.
The exceptionally stable atmospheric conditions allow for deep CMB observations during the majority of the year.
With a small-aperture, on-axis refractive design, BICEP receivers search for primordial $B$~modes at degree angular scales - the predicted peak of the $B$~mode power spectrum \cite{kamionkowski2016quest}.
Both BICEP3 and BA use transition-edge sensor (TES) bolometers cooled to 250 mK as detectors and optical elements that are cooled down to cryogenic temperatures to minimize internal load. 
To measure polarization with high accuracy, TES are coupled to a dual-slot antenna array, effectively arranging them in orthogonal and co-located detector pairs. In addition, to minimize beam systematics for pixels near the edge of the supporting frame that houses the detectors, the corrugated frame design is used \cite{soliman2018design}.
In this paper, we will refer to the orthogonal detectors in each pair as V (for vertical) and H (for horizontal) detectors (where $V \bot H$). 
BICEP3 and BA are mounted on three-axis mounts allowing movement in azimuth, elevation, and boresight angle (illustrated in Figure \ref{fig:boresight}), since observations at different boresight angles are essential in reconstructing the polarized sky signal.

To get a sensitive measurement of the tensor-to-scalar ratio, control of instrumental systematics is extremely important. 
For this reason, every austral summer we perform extensive calibration measurements on our receivers that help understand instrumental response.
One of the leading instrumental systematic is caused by beam mismatch between detectors in the same pair, resulting in some of the temperature signal leaking into polarization ($T\rightarrow{}P$ leakage) when taking the pair-difference signal. 
To understand how this affects CMB data we perform far-field beam mapping (FFBM) measurements on all of our receivers. 
This process yields high-fidelity measurements of the unpolarized response of our detectors.
In addition to FFBM data, we take measurements in different regimes and with different sources to measure sidelobes as well as polarized beam response.

Results presented in this work include data taken between 2017 and 2023, with BICEP3 which has 2400 detectors at 95~GHz, BA30/40 which has 726 detectors split between 30~GHz and 40~GHz, and BA150 which had 3,240 detectors at 150~GHz in 2023. 
BA150 will reach a fully populated focal plane unit (FPU) during the 2024/2025 austral summer, housing a total of 7776 optically-coupled detectors.
The deployment of a third BA receiver at 220/270~GHz (BA220/270) is also expected during the 2024/2025 austral summer \cite{yuka_spie}.
In section \ref{FFBM}, we present the latest FFBM measurements from previously deployed and newly deployed receivers.
In section \ref{FFBM_pol}, we detail the measurements  and analysis process of the polarized beam response of BICEP3, which we use to get an independent measurement of polarization efficiency for this receiver.
In section \ref{SL}, we present measurements of sidelobe response for BICEP3 and BA30/40, including forebaffle performance for BICEP3.

\section{Far Field Beam Measurements with Thermal Source}
\label{FFBM}

\begin{figure}[ht!]
\centering
\includegraphics[width=15cm]{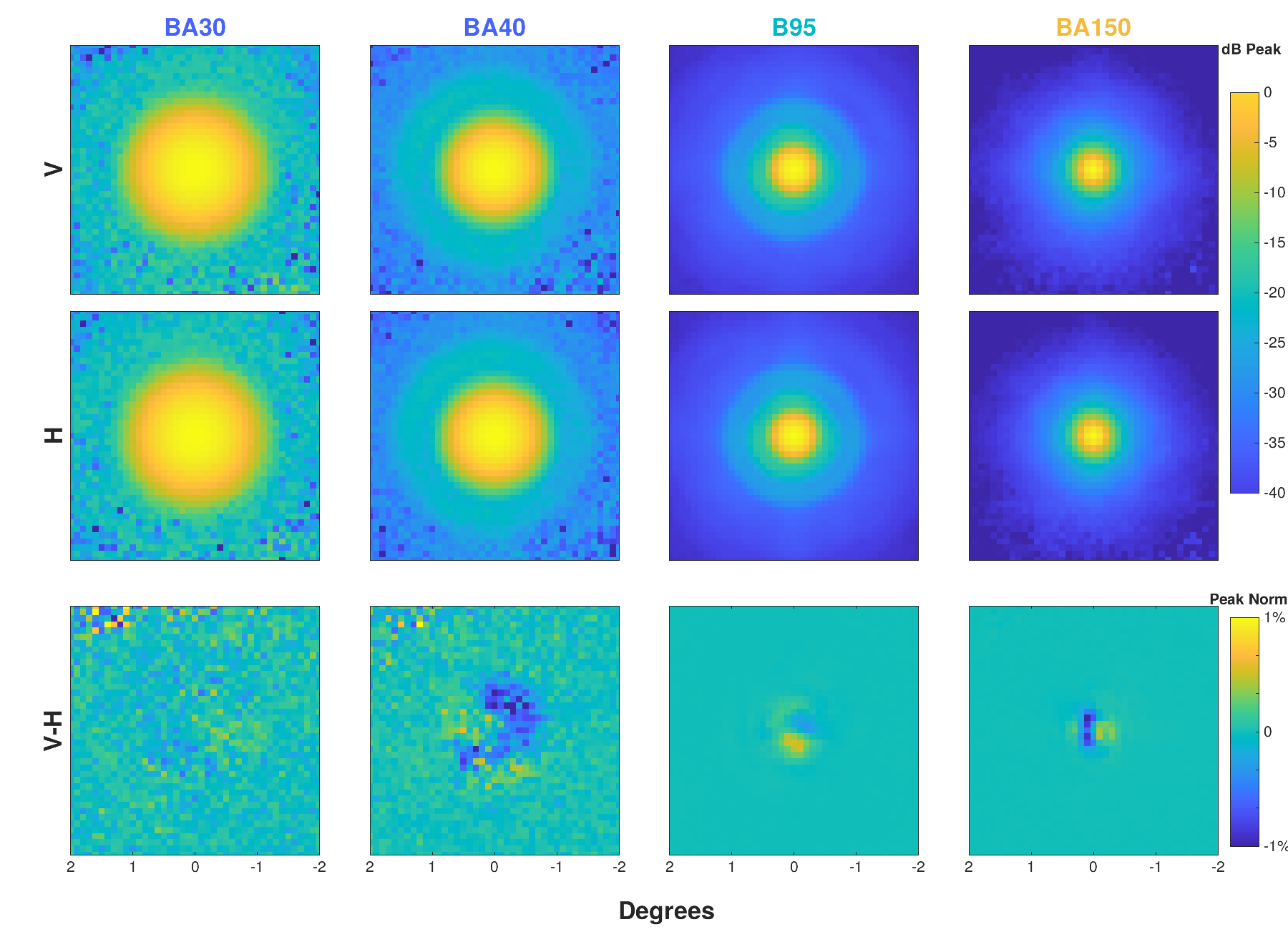}
\caption{All-detector averaged beams for BICEP3 (B95) and BICEP Array (BAXXX where XXX denotes different frequencies) instruments during austral summer 2023. From left to right, columns one and two correspond to BA30/40, the third column to  BICEP3 95~GHz, and fourth column to BA150. 1st row: V polarization beam. 2nd row: H polarization beam. 3rd row: difference (V - H) beam. Thumbnails of the main beam are plotted (-2° to 2° square axis).}
\label{fig:FFBM_updated.png}
\end{figure}

For BICEP3 and deployed BA receivers, the far-field distance ($2D^{2}/\lambda$) is shorter than 200m, except for BA150 which has a far field distance of $\sim$250 m. 
We take advantage of the locations of the Dark Sector Laboratory (DSL) that houses BICEP3 and Martin A. Pomerantz Observatory (MAPO) that houses BA, where the two buildings are about $\sim$ 200m apart to perform ground based calibration measurements. We place a mechanically chopped thermal source atop a mast on each building, so that the telescope on the opposite building can observe it.
Because our telescopes cannot point below $~45^\circ$ in elevation, we install a large, flat aluminum mirror mounted on strut legs above each receiver, to redirect its rays on the horizon.
With this setup, we scan across the source to produce a map of the beam pattern for each detector at multiple boresight orientations. Since the source is unpolarized, both V and H polarization detectors will be sensitive to the source at any time. We demodulate each detector's timestream and bin into 0.1° square pixel per-detector maps plotted in a detector-centered coordinate system defined as
$[\text{centroid}(\text{V})+\text{centroid}(\text{H})]/2$. This detector-centered coordinate system is called $\text{x}^\prime$/$\text{y}^\prime$\cite{ade2019bicep2}. We accumulate individual maps until every detector has been mapped several times at multiple boresight angles for consistency checks. 

Analysis of these data involves extracting per-detector beam parameters by fitting a 2D elliptical Gaussian described by:

\begin{equation}
    B(\textbf{x}) = \frac{1}{\Omega}\text{exp}\left[ -(\textbf{x} - \boldsymbol{\mu})^{T}
	  \mathbf{\Sigma}^{-1} (\mathbf{x} - \boldsymbol{\mu})/2 \right] + b
\end{equation}

\noindent where \(\boldsymbol{x}\) is a 2D position vector
$(x',y')$, \(\boldsymbol{\mu}\) is the beam center $(x_0,y_0)$ , \(\boldsymbol{\Omega}\) is the normalization constant, \(b\) is a background offset, and \(\boldsymbol{\Sigma}\) is the a covariance matrix defined as:
	  \begin{equation}
	      \boldsymbol{\Sigma} = \boldsymbol{R}^{-1}\boldsymbol{C}\boldsymbol{R}
	  \end{equation}

\noindent where \(\boldsymbol{R}\) is a 2x2 rotation matrix parameterized by the angle \(\gamma\) of the major axis of the elliptical Gaussian with respect to $x^\prime$ counter-clockwise, and the covariance matrix  $\boldsymbol{C}$ is given by:
	
      \begin{equation}
          \boldsymbol{C} 
	  =
	  \begin{pmatrix}
	  \sigma^{2}_{maj}  & 0 \\
	  
	  0 & \sigma^{2}_{min}
	  
	  \end{pmatrix}
      \end{equation}
	  
\noindent The 7-parameter fit consists of $[x_0,y_0,\sigma_{min},\sigma_{maj}, \gamma,\Omega,b]$, which fully characterizes the beam. 

For each detector and pair, we use the fitted parameters to derive total ellipticity $e = (\sigma_{maj}^{2}-\sigma_{min}^{2})/(\sigma_{maj}^{2}~+~\sigma_{maj}^{2})$, ellipticity plus $p = e\cos{2\gamma}$, ellipticity cross $c = e\sin{2\gamma}$, and beamwidth $\sigma = (\sigma_{maj}+\sigma_{min})/2$. 
We take the median across all derived parameters as the best estimate of each parameter and take half the width of the central 68$\%$ of the distribution of those measurements as the measurement uncertainty. 
This procedure mitigates the impact of outliers on our final statistics \cite{ade2019bicep2}. Table \ref{param_table} shows measured beam parameters from 2019 to 2024 for BA30/40 and B95. Parameters are presented as FPU median ± FPU scatter ± measurement uncertainty. Even though the BA30/40 FPU has changed between 2020 and 2024 with some 30/40~GHz detector tiles swapped, the variation of beam parameters remains within the uncertainty range. 
Comparing the B95 beam parameters from 2019 to 2023 the only significant change is a coherent decrease in beamwidth of 1.2\%. We might expect this effect given the replacement of the slap-type window to a thin window in December of 2022. The old slab-type window acted like a meniscus lens on the beam when the receiver was under vacuum. The thin window reduces the impact of the window on the focus of the instrument\cite{2022SPIE12190E..2LE} resulting in the measured beamwidth decrease.

At the end of each calibration season, we use the derived beam parameters to apply automatic cuts to the data. We then coadd the per-detector maps across all different boresight observations. 
Figure \ref{fig:FFBM_updated.png} shows array coadded beams for 4 different observing frequencies. These coadded beams are used to calculate beam window functions which are in turn used to smooth our CMB maps as described in previous work. \cite{st2020optical}

\begin{center}
\begin{table}[]
\footnotesize
    \centering
    \begin{tabular}{ |c|c|c|c| } 
        \hline
        \textbf{Parameter} & \textbf{BA30} & \textbf{BA40} & \textbf{B95} \\
        \Xhline{3\arrayrulewidth}
        \multirow{5}{3em}{$\sigma (^{\circ})$} 
         & \cellcolor[RGB]{169, 195, 182} -
        & \cellcolor[RGB]{169, 195, 182} -
        & \cellcolor[RGB]{169, 195, 182} 0.166 ± 0.004 ± 0.002\\
        & \cellcolor[RGB]{206, 223, 223} 0.478 ± 0.018 ± 0.015 
        & \cellcolor[RGB]{206, 223, 223} 0.364 ± 0.017 ± 0.009 
        & \cellcolor[RGB]{206, 223, 223} -\\ 
        & \cellcolor[RGB]{183, 209, 211} - 
        & \cellcolor[RGB]{183, 209, 211} - 
        & \cellcolor[RGB]{183, 209, 211} 0.165 ± 0.004 ± 0.002\\
        & \cellcolor[RGB]{166, 195, 206} 0.488 ± 0.024 ± 0.012
        & \cellcolor[RGB]{166, 195, 206} 0.355 ± 0.013 ± 0.007
        & \cellcolor[RGB]{166, 195, 206} 0.162 ± 0.003 ± 0.002 \\
        & \cellcolor[RGB]{143, 184, 202} 0.478 ± 0.022 ± 0.011  
        & \cellcolor[RGB]{143, 184, 202} 0.357 ± 0.013 ± 0.008
        & \cellcolor[RGB]{143, 184, 202} - \\ 
       
        \hline
        \multirow{5}{3em}{p} 
        & \cellcolor[RGB]{169, 195, 182} -
        & \cellcolor[RGB]{169, 195, 182} -
        & \cellcolor[RGB]{169, 195, 182} -0.023 ± 0.026 ± 0.018\\
        & \cellcolor[RGB]{206, 223, 223} 0.027 ± 0.046 ± 0.043 
        & \cellcolor[RGB]{206, 223, 223} 0.029 ± 0.069 ± 0.028 
        & \cellcolor[RGB]{206, 223, 223} -\\ 
        & \cellcolor[RGB]{183, 209, 211} - 
        & \cellcolor[RGB]{183, 209, 211} - 
        & \cellcolor[RGB]{183, 209, 211} 0.013 ± 0.025 ± 0.023\\
        & \cellcolor[RGB]{166, 195, 206} 0.015 ± 0.072 ± 0.035
        & \cellcolor[RGB]{166, 195, 206} 0.022 ± 0.053 ± 0.028
        & \cellcolor[RGB]{166, 195, 206} 0.010 ± 0.029 ± 0.020 \\
        & \cellcolor[RGB]{143, 184, 202} 0.009 ± 0.066 ± 0.049  
        & \cellcolor[RGB]{143, 184, 202} 0.018 ± 0.062 ± 0.040 
        & \cellcolor[RGB]{143, 184, 202} - \\ 
        \hline
        \multirow{5}{3em}{c} 
        & \cellcolor[RGB]{169, 195, 182} -
        & \cellcolor[RGB]{169, 195, 182} -
        & \cellcolor[RGB]{169, 195, 182} -0.026 ± 0.031 ± 0.017\\
        & \cellcolor[RGB]{206, 223, 223} -0.015 ± 0.035 ± 0.022 
        & \cellcolor[RGB]{206, 223, 223} 0.007 ± 0.030 ± 0.036 
        & \cellcolor[RGB]{206, 223, 223} -\\ 
        & \cellcolor[RGB]{183, 209, 211} - 
        & \cellcolor[RGB]{183, 209, 211} - 
        & \cellcolor[RGB]{183, 209, 211} -0.020 ± 0.023 ± 0.018\\
        & \cellcolor[RGB]{166, 195, 206} -0.003 ± 0.019 ± 0.036
        & \cellcolor[RGB]{166, 195, 206} -0.010 ± 0.029 ± 0.029
        & \cellcolor[RGB]{166, 195, 206} -0.009 ± 0.026 ± 0.024\\
        & \cellcolor[RGB]{143, 184, 202} 0.008 ± 0.016 ± 0.036  
        & \cellcolor[RGB]{143, 184, 202} 0.004 ± 0.027 ± 0.034 
        & \cellcolor[RGB]{143, 184, 202} - \\ 
        \hline
        \multirow{5}{3em}{$\text{d}\sigma (^{\circ})$} 
        & \cellcolor[RGB]{169, 195, 182} -
        & \cellcolor[RGB]{169, 195, 182} -
        & \cellcolor[RGB]{169, 195, 182} 2.4e-04 ± 0.001 ± 0.000\\
        & \cellcolor[RGB]{206, 223, 223} 0.003 ± 0.010 ± 0.008 
        & \cellcolor[RGB]{206, 223, 223} 1.3e-4 ± 0.008 ± 0.003 
        & \cellcolor[RGB]{206, 223, 223} -\\ 
        & \cellcolor[RGB]{183, 209, 211} - 
        & \cellcolor[RGB]{183, 209, 211} - 
        & \cellcolor[RGB]{183, 209, 211} -4.8e-04 ± 0.001 ± 0.000\\
        & \cellcolor[RGB]{166, 195, 206} 0.003 ± 0.007 ± 0.005
        & \cellcolor[RGB]{166, 195, 206} -1.9e-4 ± 0.007 ± 0.002
        & \cellcolor[RGB]{166, 195, 206} -5.0e-04 ± 0.001 ± 0.000 \\
        & \cellcolor[RGB]{143, 184, 202} 1.6e-03 ± 0.014 ± 0.006  
        & \cellcolor[RGB]{143, 184, 202} -8.5e-04 ± 0.009 ± 0.002 
        & \cellcolor[RGB]{143, 184, 202} - \\ 
        \hline
        \multirow{5}{3em}{\text{d}p} 
        & \cellcolor[RGB]{169, 195, 182} -
        & \cellcolor[RGB]{169, 195, 182} -
        & \cellcolor[RGB]{169, 195, 182} 0.004 ± 0.012 ± 0.002\\
        & \cellcolor[RGB]{206, 223, 223} 0.014 ± 0.018 ± 0.019 
        & \cellcolor[RGB]{206, 223, 223} 0.014 ± 0.018 ± 0.019 
        & \cellcolor[RGB]{206, 223, 223} -\\ 
        & \cellcolor[RGB]{183, 209, 211} - 
        & \cellcolor[RGB]{183, 209, 211} - 
        & \cellcolor[RGB]{183, 209, 211} -0.003 ± 0.012 ± 0.003\\
        & \cellcolor[RGB]{166, 195, 206} 0.001 ± 0.027 ± 0.020
        & \cellcolor[RGB]{166, 195, 206} -0.001 ± 0.031 ± 0.009
        & \cellcolor[RGB]{166, 195, 206} -0.004 ± 0.013 ± 0.002 \\
        & \cellcolor[RGB]{143, 184, 202} 0.006 ± 0.034 ± 0.021  
        & \cellcolor[RGB]{143, 184, 202} 0.001 ± 0.021 ± 0.013
        & \cellcolor[RGB]{143, 184, 202} - \\ 
        \hline
        \multirow{5}{3em}{\text{d}c} 
        & \cellcolor[RGB]{169, 195, 182} -
        & \cellcolor[RGB]{169, 195, 182} -
        & \cellcolor[RGB]{169, 195, 182} -0.001 ± 0.004 ± 0.002\\
        & \cellcolor[RGB]{206, 223, 223} 0.006 ± 0.013 ± 0.013 
        & \cellcolor[RGB]{206, 223, 223} 0.001 ± 0.015 ± 0.010 
        & \cellcolor[RGB]{206, 223, 223} -\\ 
        & \cellcolor[RGB]{183, 209, 211} - 
        & \cellcolor[RGB]{183, 209, 211} - 
        & \cellcolor[RGB]{183, 209, 211} -0.002 ± 0.004 ± 0.002\\
        & \cellcolor[RGB]{166, 195, 206} -0.001 ± 0.017 ± 0.014
        & \cellcolor[RGB]{166, 195, 206} 0.001 ± 0.010 ± 0.009
        & \cellcolor[RGB]{166, 195, 206} -0.002 ± 0.004 ± 0.004 \\
        & \cellcolor[RGB]{143, 184, 202} 0.009 ± 0.012 ± 0.018  
        & \cellcolor[RGB]{143, 184, 202} 0.004 ± 0.014 ± 0.011 
        & \cellcolor[RGB]{143, 184, 202} - \\ 
        \hline
        \multirow{5}{3em}{$\text{dx}(\prime)$} 
        & \cellcolor[RGB]{169, 195, 182} -
        & \cellcolor[RGB]{169, 195, 182} -
        & \cellcolor[RGB]{169, 195, 182} -0.05 ± 0.14 ± 0.05\\
        & \cellcolor[RGB]{206, 223, 223} 0.71 ± 1.55 ± 0.52 
        & \cellcolor[RGB]{206, 223, 223} -0.51 ± 1.91 ± 0.35 
        & \cellcolor[RGB]{206, 223, 223} -\\ 
        & \cellcolor[RGB]{183, 209, 211} - 
        & \cellcolor[RGB]{183, 209, 211} - 
        & \cellcolor[RGB]{183, 209, 211} -0.06 ± 0.15 ± 0.04\\
        & \cellcolor[RGB]{166, 195, 206} 0.58± 2.67 ±0.60
        & \cellcolor[RGB]{166, 195, 206} 0.02± 1.40±0.45
        & \cellcolor[RGB]{166, 195, 206} -0.09 ± 0.14 ± 0.06 \\
        & \cellcolor[RGB]{143, 184, 202} 1.55 ± 2.10 ± 0.86  
        & \cellcolor[RGB]{143, 184, 202} -0.21 ± 1.51 ± 0.60 
        & \cellcolor[RGB]{143, 184, 202} - \\ 
        \hline
        \multirow{5}{3em}{$\text{dy}(\prime)$} 
        & \cellcolor[RGB]{169, 195, 182} -
        & \cellcolor[RGB]{169, 195, 182} -
        & \cellcolor[RGB]{169, 195, 182} 0.02 ± 0.16 ± 0.06\\
        & \cellcolor[RGB]{206, 223, 223} -0.21 ± 1.37 ± 0.39 
        & \cellcolor[RGB]{206, 223, 223} -0.07 ± 0.57 ± 0.32 
        & \cellcolor[RGB]{206, 223, 223} -\\ 
        & \cellcolor[RGB]{183, 209, 211} - 
        & \cellcolor[RGB]{183, 209, 211} - 
        & \cellcolor[RGB]{183, 209, 211} -0.01 ± 0.14 ± 0.06\\
        & \cellcolor[RGB]{166, 195, 206} 0.19± 0.78± 0.57
        & \cellcolor[RGB]{166, 195, 206} -0.43± 0.81± 0.44
        & \cellcolor[RGB]{166, 195, 206} 0.01 ± 0.15 ± 0.06 \\
        & \cellcolor[RGB]{143, 184, 202} 0.47 ± 1.03 ± 0.50  
        & \cellcolor[RGB]{143, 184, 202} -0.06 ± 0.59 ± 0.44 
        & \cellcolor[RGB]{143, 184, 202} - \\ 
        \hline
    \end{tabular}
    \caption{Beam parameter summary table for BA30/40 and B95 for years \colorbox[RGB]{169, 195, 182}{2019}, \colorbox[RGB]{206, 223, 223}{2020}, \colorbox[RGB]{183, 209, 211}{2021}, \colorbox[RGB]{166, 195, 206}{2023}, and \colorbox[RGB]{143, 184, 202}{2024}. The values listed as FPU median ± FPU scatter ± measurement uncertainty. The variation of each receiver from year to year is smaller than the measurement uncertainty, except from the beamwidth parameter of the B95 receiver. This coherent shift of $\sim1.2\%$ is due to the thin window change on December 2022\cite{2022SPIE12190E..2LE}. All beamwidth parameters have been corrected for the non-negligible size of the chopped source aperture in the far field.}
    \label{param_table}
\end{table}
\end{center}

Due to beam mismatch within a detector pair, residual beam response in the pair-difference beams will cause $T\rightarrow{}P$ leakage. 
We deproject to filter out the leading-order terms of this leakage as described in, e.g., BK-III \cite{ade2015bicep2}. 
Any power remaining in the difference beam response after deprojection contributes to so-called undeprojected $T\rightarrow{}P$ leakage. 
We want to fully characterize this type of systematic error. 
To do so we estimate the bias on $r$ resulting from it by using specialized beam simulations and combining their output as part of our multicomponent likelihood analysis\cite{ade2019bicep2,ade2021improved}.
Figure \ref{fig:FFBM_updated.png} shows the pair-difference beams; the undeprojected residual beams are work-in-progress and will be shown in a future publication, as well as the expected impact on CMB measurements and inflation constraints (bias on $r$). 

\section{Far Field Beam Measurements with Polarized Source at 95 GHz}
\label{FFBM_pol}

A subdominant instrumental systematic that arises due to per-pair polarization mismatch is mixing of polarization modes, in this case $E\rightarrow{}B$ leakage. 
Since the amount of $E$-mode power is much lower than temperature at any given angular scale we expect the effect to be much less prominent than $T\rightarrow{}P$ leakage.
However, understanding the amount of $E\rightarrow{}B$ leakage for the BICEP3 receiver helps us to fully characterize its performance,
and even a small effect might become relevant in the context of future, more sensitive CMB experiments.

\subsection{Measurement setup}
To measure the polarized beam response, we utilize the same setup as described in the previous section (combination of source on the mast + mirror), but instead of using a thermal source we use a linearly polarized source. The source is described in detail in previous work;\cite{cornelison2022improved} here we give a brief overview.
The source is a broad spectrum noise source (BSNS) with a linearly polarized output waveguide equipped with a wire grid polarizer. 
It is kept thermally stable under a PID feedback loop and electronically chopped at 16 Hz. 
The source emission band center is at 95~GHz, and peak power can be controlled by a set of dialectric waveguide attenuators. 
For this measurement the wire grid is aligned with gravity and not allowed to rotate. 
We deploy the source on top of a 12-meter tall mast on MAPO, 200 meters away from BICEP3 and scan each detector across the source at multiple boresight (referred to as ``DK'') angles.

\begin{figure}[ht!]
\centering
\includegraphics[width=16cm]{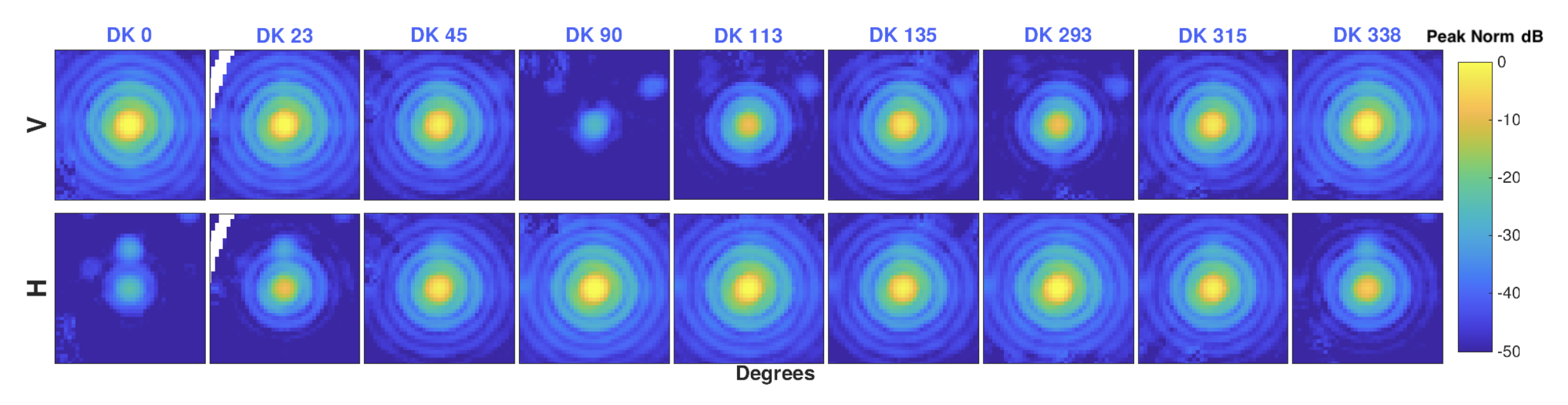}
\caption{Beam maps of a single 95~GHz BICEP3 pair plotted in a peak normalized logarithmic scale from -2° to 2° using 0.1° pixels. First row: Polarization V across our different boresight angles. Second row: the corresponding orthogonal pair. Due to the BSNS brightness the Airy ring patterns that arise from truncating the beam at the aperture are clearly visible. The beam response is modulated as a function of boresight rotation of BICEP3 with peak beam response having a 90° offset between V and H detectors. The secondary peaks at $\sim$ -25 dB that are roughly 1 degree from the main beam are due to crosstalk in the time-division readout system and have been previously characterized \cite{ade2015bicep2}. }
\label{fig:bsns_components.png}
\end{figure}

The mapmaking procedure is the same as with FFBM. Instead of relying on pair-differencing to reconstruct the polarization signal of the source, we decouple temperature from polarization by using the maps at the different boresight angles. The per-detector maps at different boresight orientations are shown in Figure \ref{fig:bsns_components.png}. The individual detector response will be modulated as a function of boresight rotation since the BSNS is linearly polarized.

\subsection{Mapmaking framework}

For an arbitrary coordinate system, the full detector response expressed in the Stokes T, Q and U basis is dominated by the field of the BSNS and given by the integral of the temperature and polarization over the detector beam. In general, each BICEP3 detector measures:

\begin{equation}
\label{full_int}
    D = 
    \int 
    \left[
    \text{B}_{\text{T}}(\mathbf{x})T+ 
    \textbf{B}^{\textbf{T}}_{\text{p}} \textbf{R} \textbf{P} 
    \right]
    \delta(\mathbf{x}-\mathbf{x^{\prime}})
    d\mathbf{x^{\prime}}
\end{equation}

\noindent where:
    
\begin{equation}
\label{full_int_def}
    \textbf{B}_{\text{p}} = 
    \begin{bmatrix}
	  \text{B}_{\text{Q}}(\mathbf{x}) \\  
	  \text{B}_{\text{U}}(\mathbf{x}) 
	\end{bmatrix}
    \text{,}\quad 
    \textbf{R} = 
     \begin{bmatrix}
	     \cos{2\phi} & -\sin{2\phi} \\
         \sin{2\phi} & \cos{2\phi} 
	 \end{bmatrix}   
    \text{, and}\quad
     \textbf{P} = 
     \begin{bmatrix}
	     \text{Q}\\  
	     \text{U} 
	 \end{bmatrix}
\end{equation}

\noindent where T is the amplitude of the unpolarized field, and Q, U the amplitudes of the two degrees of freedom of the polarized field of the source. The temperature and polarization beam responses of a given detector are denoted by \(B_{T}\), \(B_{Q}\), and \(B_{U}\). In addition, $\phi$ is the angle between the co-polar axis of each detector and that of the source. Our calibration source has an aperture of a few centimeters and is located in the far field ($\sim$ 200m away), therefore it can be approximated as a point source located at \(\mathbf{x^{\prime}}\).

\noindent Based on this definition, since the source is linearly polarized we expect our detectors to have maximum beam response in Stokes Q and minimum beam response in Stokes U. Beam response in the Stokes U map would represent residual power that would mix $E$-modes with B-modes. Therefore, the total power in the Stokes U map should integrate to zero. With this setup, we can solve for \(B_{T}\), \(B_{Q}\), and \(B_{U}\) for each individual pixel in the resulting map.

A single polarized detector measures the average power along a particular axis, which is oriented at an angle $\phi$ relative to Q (the polarization axis of the source). Using the Stokes basis it measures:

\begin{equation}
   d = B_{T} + B_{Q}\cos{2\phi} + B_{U}\sin{2\phi} +n
\end{equation}

\noindent where n is the detector noise, which we assume to be stationary with zero mean and variance \( \sigma^{2}\). We take many per detector observations, \( d_{i} \), at a range of angles \( \phi_{i} \), weighted by $w_{i}$. Given the stability of the environment at the South Pole and that the source housing is thermally controlled, we apply uniform weighting of the data. Summing all the per-pixel samples gives:  

\begin{align}
    z &= \sum_i w_{i}d_{i} \notag \\
    \left\langle z \right \rangle &= {B_{T}}\sum_i w_{i} + {B_{Q}} \sum_i w_{i}\cos{2\phi_{i}} + {B_{U}} \sum_i w_{i}\sin{2\phi_{i}}
\end{align}

\noindent We have used our previous assumption that \( \left\langle n \right\rangle = 0 \). 
We can now calculate a few similar quantities:

\begin{align}
    x &= \sum_i d_{i} w_{i}\cos{2\phi_{i}} \notag \\
    \left\langle x \right \rangle 
    &= {B_{T}} \sum_i w_{i}\cos{2\phi_{i}} + {B_{Q}} \sum_i w_{i}\cos^{2}{2\phi_{i}} + {B_{U}} \sum_i w_{i}\sin{2\phi_{i}}\cos{2\phi_{i}}
\end{align}

\noindent Similarly:

\begin{align}
    y &= \sum_i d_{i} w_{i}\sin{2\phi_{i}} \notag \\
    \left\langle y \right \rangle 
    &= {B_{T}} \sum_i w_{i}\sin{2\phi_{i}} + {B_{U}} \sum_i w_{i}\sin^{2}{2\phi_{i}} + {B_{Q}} \sum_i w_{i}\sin{2\phi_{i}}\cos{2\phi_{i}}
\end{align}

\noindent Where \( \left\langle  x \right\rangle \) is the time averaged Stokes Q component of the signal measured by our detector and \( \left\langle  y \right\rangle \) is the time averaged signal measured perpendicular to Q. \newline

\noindent Combining the above equations we can write them in matrix form as follows:

\begin{equation}
\label{matrix}
\centering
\begin{bmatrix}
    \left\langle z \right\rangle  \\
    \left\langle x \right\rangle  \\
    \left\langle y \right\rangle  \\
\end{bmatrix}
=
\begin{bmatrix}
    \sum w_{i}  & \sum w_{i}\cos{2\phi_{i}}  & \sum w_{i}\sin{2\phi_{i}} \\
    
    \sum w_{i}\cos{2\phi_{i}} & \sum w_{i}\cos^{2}{2\phi_{i}} & \sum w_{i}\sin{2\phi_{i}}\cos{2\phi_{i}} \\
    
    \sum w_{i}\sin{2\phi_{i}} & \sum w_{i}\sin{2\phi_{i}}\cos{2\phi_{i}} & \sum w_{i}\sin^{2}{2\phi_{i}} 
\end{bmatrix}
\space
\begin{bmatrix}
    \hat{B_{T}}\\
    \hat{B_{Q}}\\
    \hat{B_{U}}
\end{bmatrix}
\end{equation}

\noindent Where  $\hat{\text{B}_{\text{T}}}$,  $\hat{\text{B}_{\text{Q}}}$,  $\hat{\text{B}_{\text{U}}}$ denote estimated quantities as opposed to the true quantities in equation \ref{full_int} and \ref{full_int_def}. We then add up the contributions to $ \left\langle z \right\rangle $, $ \left\langle x \right\rangle  $, and $ \left\langle y \right\rangle  $ along with the various terms in the matrix from equation \ref{matrix}. Then invert the resulting matrix to solve for \(\hat{B_{T}}\), \(\hat{B_{Q}}\), and \(\hat{B_{U}}\). It is worth noting that we need at least two different boresight angles of observation to be able to invert the matrix in equation \ref{matrix}. In the limit of infinite and distinct boresight measurements we fully reconstruct the true polarization response of the source. The total number of observations we have are 13 at 9 different boresight angles. Given that our boresight angle coverage is limited, to understand how well each pixel is measured we compute a simple covariance matrix using \( \text{Var} ( X) = \left\langle X^{2} \right\rangle -  \left\langle X \right\rangle^{2}  \), as well as \( \text{Cov} ( X,Y) = \left\langle X Y \right\rangle -  \left\langle X \right\rangle  \left\langle Y \right\rangle \), where \( \text{X} \) and \( \text{Y} \) represent \(\hat{B_{T}}\), \(\hat{B_{Q}}\), or \(\hat{B_{U}}\) and their respective measurement averages.

\begin{figure}[ht!]
\centering
\includegraphics[width=15cm]{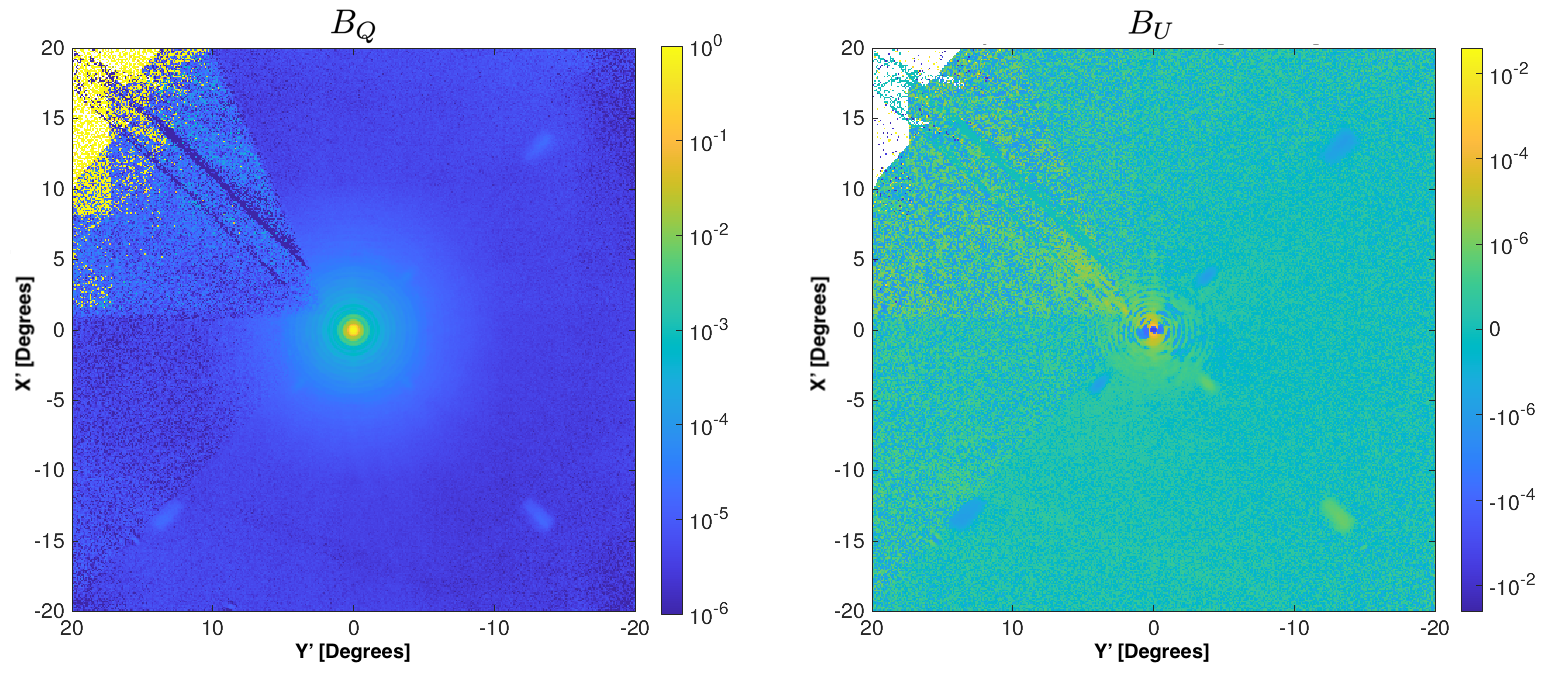}
\caption{BICEP3 polarized beam response to linearly polarized source in the far field. All-detector coadded $\text{B}_{\text{Q}}$ (left), all-detector coadded $\text{B}_{\text{U}}$ map plotted using negative logarithmic plot with zero-point set to 10e-7 (right). The colorbar is normalized to the peak beam response in $\text{B}_{\text{T}}$. More than 1900 detectors contribute to these coadded maps out of 2200, hosted in BICEP3's focal plane. The top left part of both maps appears much noisier than the rest of the maps and is due to the case of insufficient $\phi$ coverage for that region of the map. Polarized features at $\sim$ 4 and $\sim$ 13 degrees  can be seen in the extended region of the coadded $\text{B}_{\text{U}}$ map.}
\label{fig:QUmaps.png}
\end{figure}

To correctly reconstruct the polarization response of the source we rely on knowing $\phi$ with very little uncertainty. Accounting for any uncertainty in $\phi$ is important because any misalignment in the detector-source orientation would cause non-zero power in the $\text{B}_{\text{U}}$ map. We refer to this effect as the first-order beam response (or ``monopole" response) leaking into the Stokes U map. This effect is not to be interpreted as the $E\rightarrow{}B$ leakage we are attempting to quantify. To estimate any uncertainty in $\phi$, we perform absolute measurements of the polarization response of our detectors during separate calibration measurements. We use a specifically designed Rotating Polarized Source (RPS) to measure the polarization response at multiple source and telescope boresight rotation angles, to fully map the detector response \cite{cornelison2022improved}. From these measurements the absolute per-detector polarization angles are derived. In this analysis, we use the RPS derived per-detector polarization angles for \(\phi\). Figure \ref{fig:QUmaps.png} shows the resulting $\text{B}_{\text{Q}}$ and $\text{B}_{\text{U}}$ polarization maps when coadded over all detectors that pass data quality cuts and were used in BK18 \cite{ade2021improved}. 

\subsection{Results}

\noindent Since we solve for our estimate of $\text{B}_{\text{T}}$, $\text{B}_{\text{Q}}$, and $\text{B}_{\text{U}}$ for every pixel in our maps, we expect that regions of insufficient $\phi$ coverage will have higher variance. An example of such a region is the top left of the maps shown in Figure \ref{fig:QUmaps.png}. In practice, we have calculated the variance of our measurement for these pixels and they will be significantly down weighted in future $E\rightarrow{}B$ analysis. The $\text{B}_{\text{U}}$ map does not show signs of any monopole beam response which gives us confidence that we have correctly accounted for any uncertainty in $\phi$ in analysis and that the RPS measurements presented in previous work \cite{cornelison2022improved} yield consistent results across different years and across different calibration measurements. However, there is still non-monopole residual power of $O(10^{-2})$ which unless accounted for would cause $E\rightarrow{}B$ leakage. The origin of the polarized features at $\sim$ 4 and $\sim$ 13 degrees is currently being investigated. 

In addition, by solving for $\text{B}_{\text{T}}$, $\text{B}_{\text{Q}}$, and $\text{B}_{\text{U}}$ per-detector, we are able to get an indirect measurement of polarization efficiency of our detectors by comparing the peak unpolarized response  with the peak polarized response. The ratio yields 99.1\(\%\) or higher for all detectors in this analysis which is consistent with the more precise measurements described in the RPS analysis \cite{cornelison2022improved}, and therefore provides an independent cross-check.

In the future, we would use the per-detector $\text{B}_{\text{U}}$ to place an upper limit on the amount of \(E\rightarrow{}B\) leakage within a pair using a process similar to beam simulations described in the previous section. 
Given that the $E$-mode power is much lower than that of temperature we expect the amount of such leakage to be much less prominent as evidenced by systematics analysis of previous experiments \cite{ade2015bicep2}.

\section{Sidelobe Measurements}

\label{SL}
Sidelobes coupling to the sky and/or the environment can be a source of increased loading and systematic errors.
Identifying such sources of contamination during CMB observations can be difficult. 
To minimize such contamination, the baffling scheme for both BICEP3 and BA consists of a large reflective ground shield and a comoving, cylindrical and blackened forebaffle at ambient temperature installed on top of the receiver window. 
This ensures that the double-diffraction criterion is respected, i.e., a ray from the ground would have to diffract twice before coupling to the receiver window.

\noindent In this section, we describe calibration measurements that directly map the sidelobes of our receivers to characterize their optical performance at a wide range of angles away from the main beam.
We will refer to these measurements as ``sidelobe beam mapping" for the rest of this paper.

\begin{figure}[hbt!]
\centering
\includegraphics[width=10cm]{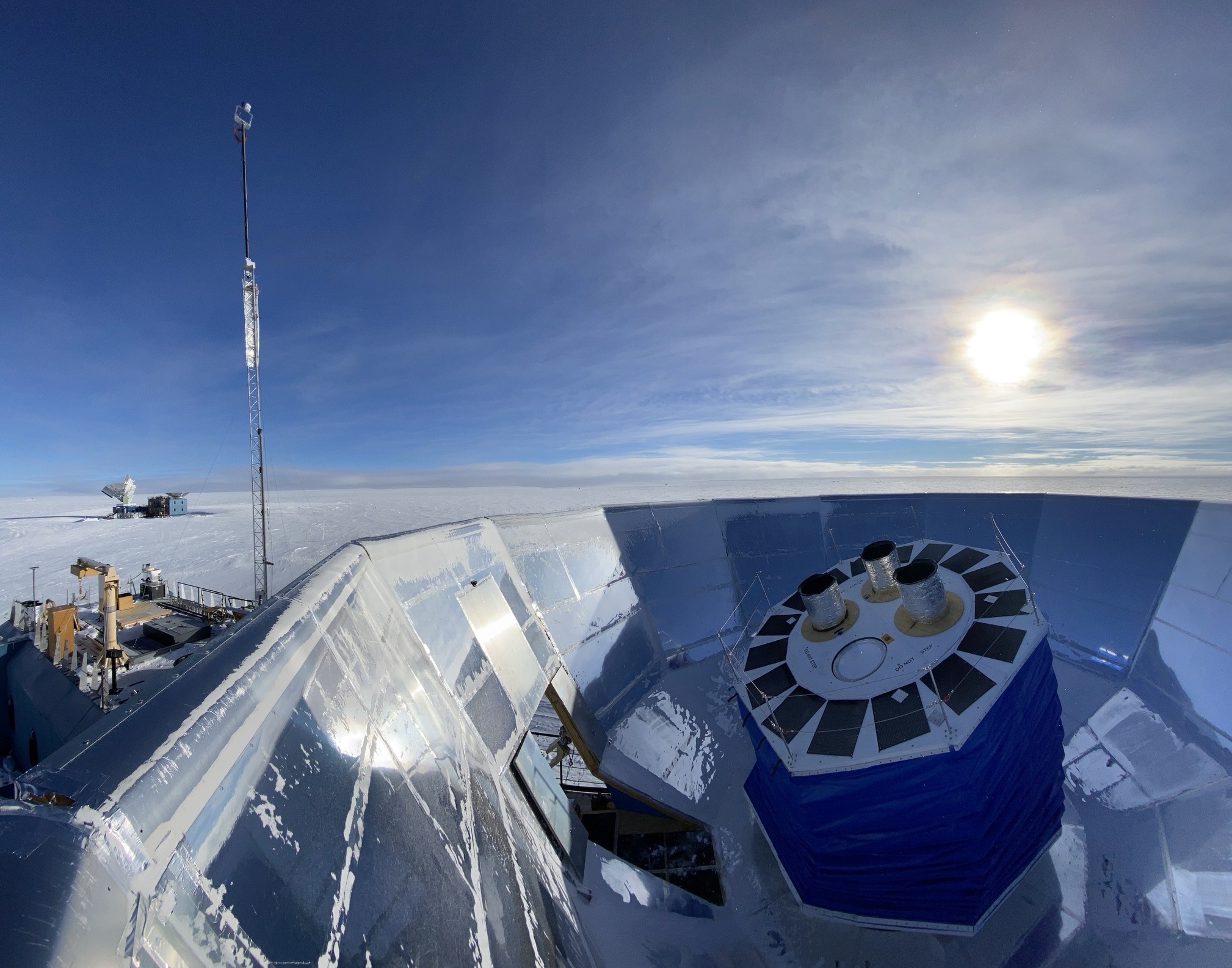}
\caption{BA30/40 performing sidelobe measurements in March 2021. BA at the time of the taking of this photo hosts three Keck Array receivers and our 30/40~GHz receiver. This receiver has the comoving forebaffle uninstalled during this schedule of sidelobe measurements while the 3 Keck array receivers have them installed. Note the 40 GHz calibration source mounted on the mast in the near field. In the distance, the building hosting BICEP3 and the South Pole Telescope (SPT) can be seen. Photo courtesy of Brandon Amat.}
\label{fig:BA1_fsl.jpg}
\end{figure}

\subsection{Measurement setup}
The measurement setup for these observations is different than in the previous sections. 
Instead of deploying the source in the far field, we place it on top of a 12-meter tall mast located on the same building as the instrument (MAPO for BA and DSL for BICEP3).
Figure \ref{fig:BA1_fsl.jpg} gives an illustration of the measurement setup taken during sidelobe measurements with our 30/40 GHz receiver during austral summer of 2020/2021. 
We use the same 95GHz source for BICEP3, and a 40~GHz source for BA30/40. The 40~GHz BSNS is very similar to the 95~GHz in construction. The key being the use of fewer frequency amplifiers in the RF chain. The band pass of the source is also sufficiently wide to cover both the 30~GHz and 40~GHz detectors. 
Once installed on the mast, we tilt the mast stage down towards the receiver and perform 380° raster scans at a fixed range of azimuth while stepping in elevation.
This time, we make full use of the attenuators installed on the source
to explore the full 70 dB dynamic range of the BSNS. 
A 90° precision twist is optionally placed before the feedhorn that allows us to take measurements at two orthogonal polarizations.

To map the extended beam regions with good  signal-to-noise, we take measurements at both source orientations, at six distinct boresight angles and at various source attenuation settings. We use an attenuation setting appropriate to each region we want to map to avoid our detectors saturating while optimizing the signal-to-noise ratio. A ``low" source power setting for the main beam region (within $\sim$2 degrees of the beam center), a ``medium" setting for the near/mid sidelobe region (defined as the region comprised between 2° and $\sim$20° and contained within the forebaffle), and a ``high" power setting for the far sidelobe region. 
The data shown in this work were taken for BICEP3 during austral summer of 2017/2018 and for BA30/40 during the austral summer of 2021/2022.

\begin{figure}[ht!]
\centering
\includegraphics[width=14cm]{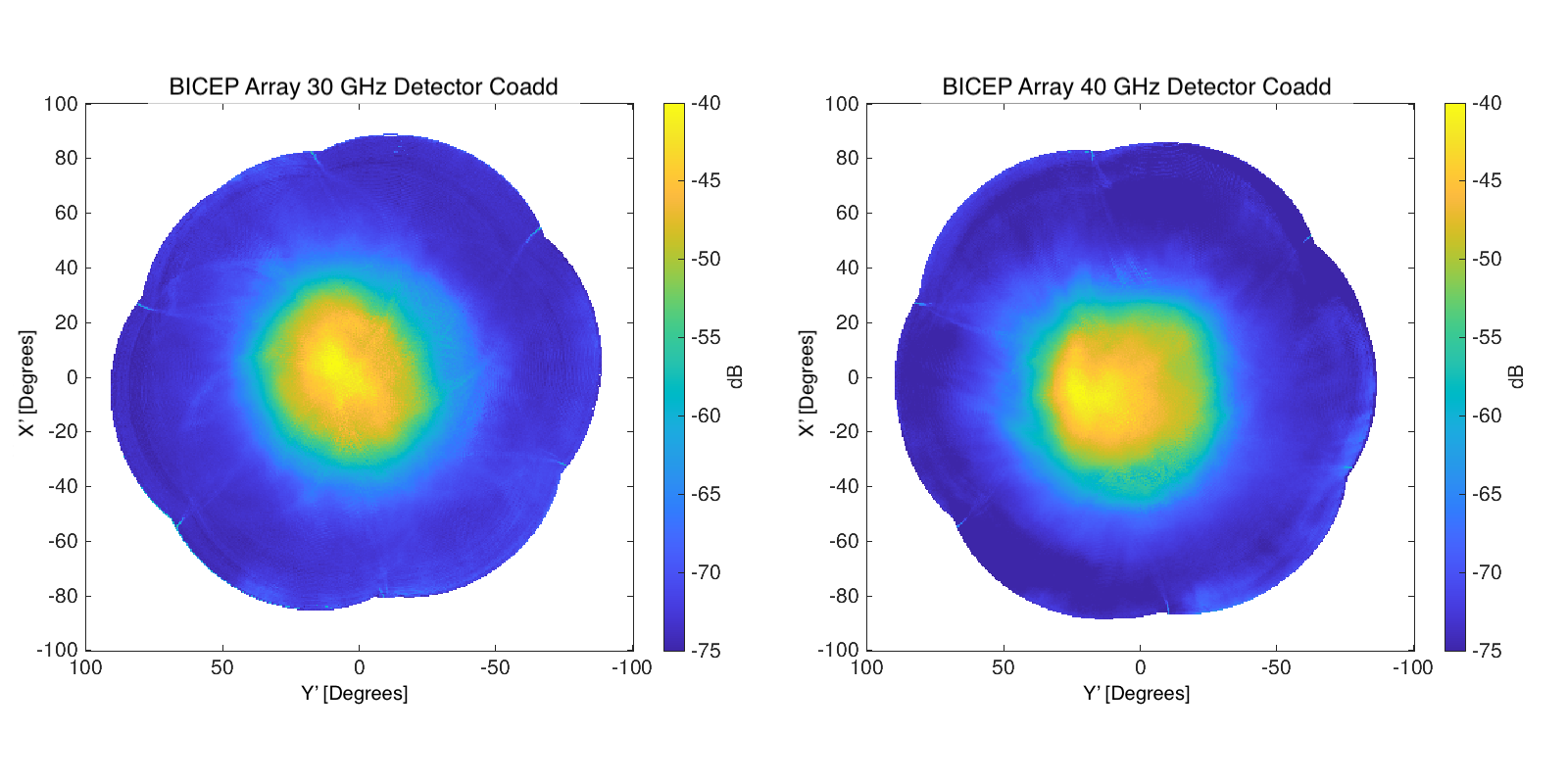}
\caption{BA30/40 sidelobe coadded maps with forebaffle installed. Left: 30~GHz coadded map, right: 40~GHz coadded map. The color axis is logarithmic and normalized to the peak response from of the FFBM beam. Note that for these measurements the main beam is not resolved due to the scan pattern. The maps show significant attenuation of power starting at \(\sim\)20° which coincides with the location of the forebaffle. The maps are normalized to the peak FFBM response which peaks roughly 40 dB higher. The stripes seen at the very edges of the maps are due to sun contamination in the data.}
\label{fig:fsl_array_coadd_30:40GHz_norm.png}
\end{figure}

\subsection{Map-making}
Mapmaking for this analysis is all done per detector. After data taking, we demodulate the time-ordered-data, calculate real pointing timestreams in $\text{x}^{\prime}$/$\text{y}^{\prime}$, and bin using 0.5° pixels into sidelobe maps for each attenuation setting. The maps at the two source polarizations are coadded to produce unpolarized maps. Then, we coadd across all boresight measurements for each power setting. For two given source power settings we take slices in $\text{x}^{\prime}$ or $\text{y}^{\prime}$ of the per detector coadded maps and regress them to get the stitch factor needed to compare the 2 power settings. The maps are then stitched across the three power settings to produce a single stitched map per detector with good signal-to-noise in all regions. It is worth noting that even with the lowest source power due to the source being only \(\sim\) 15 meters away, detectors show signs of saturation in the main beam region.
For the BA30/40 the main beam is not resolved at all in the sidelobe measurements (partly due to the observation pattern not scanning directly across the source). Figure \ref{fig:fsl_array_coadd_30:40GHz_norm.png} shows the resulting maps, which are normalized to the peak beam response of the beam measured during FFBM.

\begin{figure}[ht!]
\centering
\includegraphics[width=15cm]{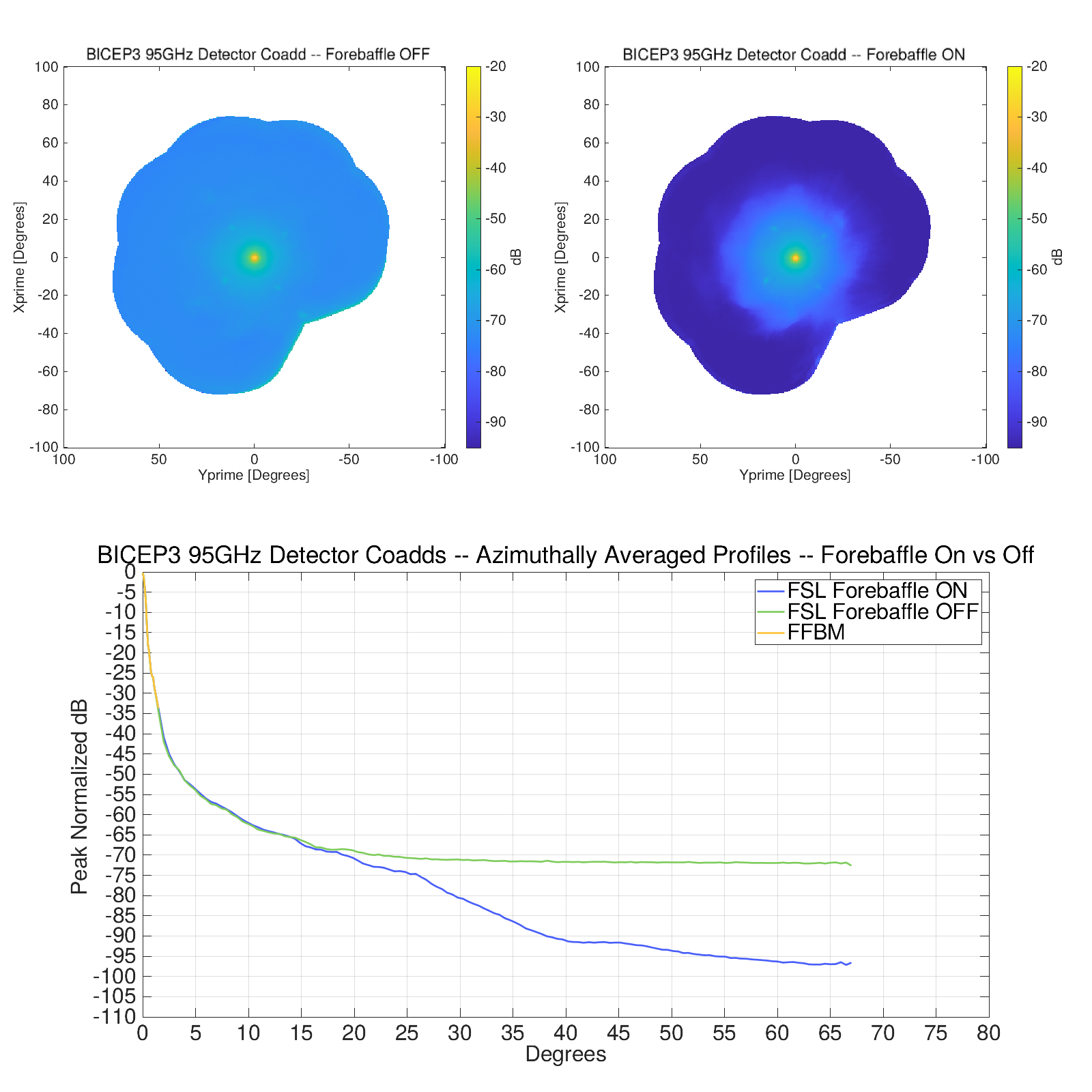}
\caption{BICEP3 95 GHz sidelobe coadded maps. Top left: all-detector coadded map with forebaffle uninstalled. Top right: all-detector coadded map with forebaffle installed. The main beam is better resolved than in Figure 7 but its amplitude is suppressed due to saturation. Bottom: azimuthally averaged profile of the stitched FFBM + sidelobe beam response. The yellow line denotes the region we use the FFBM coadded profiles. Y-axis is logarithmic and normalized to the fitted peak of the FFBM main beam. The profiles show a significant attenuation starting at \(\sim\)18° which coincides with the location of the forebaffle for BICEP3. The attenuation level between the two forebaffle configurations is \(\sim\)25 dB.}
\label{fig:fsl_b3_array_coadd_fbon:off_norm.png}
\end{figure}

\subsection{Forebaffle performance for BICEP3}
Figure \ref{fig:fsl_b3_array_coadd_fbon:off_norm.png} shows the corresponding coadded maps for BICEP3 95 GHz. For this dataset we have complete measurements to make coadded sidelobe maps with and without the forebaffle installed. This allows us to visualize the forebaffle performance in both map space and in the azimuthally averaged beam profile. To increase the fidelity of the main beam relative to the sidelobe region, we use the FFBM beam for the first 2 degrees in the radial profiles and normalize relative to the fitted peak of the FFBM beam. As expected the sidelobe power is rapidly attenuated starting at \(\sim\)18° away from the main beam because of the forebaffle.
The difference between forebaffle ON and OFF is clearly visible on Figure \ref{fig:fsl_b3_array_coadd_fbon:off_norm.png}.
This effect, although smoothed by the azimuthal average, is captured in the radial profiles. This results in a -25 dB attenuation in power, illustrating the forebaffle is absorbing sidelobe power as intended. In addition, the polarized features also seen in the far field maps in the previous section, can be seen here as well at the same radius of \(\sim\)14° away from the main beam. After performing an elliptical Gaussian fit to them, similar to that described in section 2, we find that their amplitude is \(\sim\) -60 dB below the peak of the main beam.  

\section{Conclusions}

In these proceedings we present analysis of FFBM data with a thermal source for BICEP3 95 GHz, BA30/40, and BA150, all taken during the austral summer of 2023. These measurements directly inform our instrumental systematics studies since we use the per-pair differential beam response to quantify the amount of \(T\rightarrow{}P\) leakage in our CMB data via beam simulations. 
We present additional calibration measurements to assess the optical performance of our polarimeters.
In particular, we presented high precision far-field polarized beam mapping with a 95 GHz linearly polarized source that in the future will be used to set an upper limit on $E\rightarrow{}B$ leakage due to non-monopole U response. 
This dataset also constitutes an independent measurement of polarization efficiency of our detectors, found to be above 99.1$\%$ for all working BICEP3 detectors. 
Such measurements have not yet been taken with BA. 
Finally, we present sidelobe measurements with both BICEP3 and BA30/40, characterizing the optical response of our receivers up to 65 degrees from the main beam.
We used these maps to measure forebaffle attenuation for BICEP3 of $-25$ dB, consistent with instrument design.

\section{Acknowledgments}
The BICEP/Keck experiments have been funded through U.S. National Science Foundation grants most recently including 2220444-2220448, 2216223, 1836010, and 1726917. The development of antenna-coupled detector technology was supported by the Jet Propulsion Laboratory Research and Technology Development fund. Focal plane development and testing were supported by the Gordon and Betty Moore Foundation at the California Institute of Technology. Readout electronics were supported by the Canada Foundation for Innovation grant to the University of British Columbia. The computations in this paper were run on the Cannon cluster supported by the FAS Science Division Research Computing Group at Harvard University. The analysis effort at Stanford University and the SLAC National Accelerator Laboratory was partially supported by the Department of Energy. We thank the staff of the U.S. Antarctic Program and in particular the South Pole Station without whose help this research would not have been possible.
We also thank our winter-over operators: Manwei Chan, Karsten Look, Calvin Tsai, Paula Crock, Ta Lee Shue, Grantland Hall, Hans Boenish, Robert Schwarz, Sam Harrison, Anthony DeCicco, Thomas Leps, Brandon Amat, Nathan Precup, Steffen Richter, Thibault Romand, and Danielle Simmons.

\bibliography{report} 
\bibliographystyle{spiebib} 

\end{document}